# On Modal Logics for Qualitative Possibility in a Fuzzy Setting

Petr Hájek
Dagmar Harmancová
Institute of Computer Science
Academy of Sciences
18207 Prague, Czech Republic
e-mail: {hajek,dasa}@uivt.cas.cz

Francesc Esteva
Pere Garcia
Lluís Godo
Artificial Intelligence Research Institute (IIIA)
Spanish Council for Scientific Research (CSIC)
e-mail: {esteva,pere,godo}@ceab.es

## Abstract

Within the possibilistic approach to uncertainty modeling, the paper presents a modal logical system to reason about qualitative (comparative) statements of the possibility (and necessity) of fuzzy propositions. We relate this qualitative modal logic to the many-valued analogues MVS5 and MVKD45 of the well known modal logics of knowledge and belief S5 and KD45 respectively. Completeness results are obtained for such logics and therefore, they extend previous existing results for qualitative possibilistic logics in the classical non-fuzzy setting.

## 1 INTRODUCTION

In the recent past, a lot of effort has been put in relating numerical and symbolic approaches to uncertain reasoning. Numerical formalisms attach degrees of belief to propositions. Belief degrees are understood as a measure on the set of possible worlds (possible state descriptions) that assigns to every proposition the measure of the set of worlds in which the proposition is true. Therefore, uncertainty measures are not truth-functional, as it is well known and established, i.e. the measure of a compound formula can not be in general obtained as a function of the measures of its subformulas.

Possibilistic logic (cf. e.g. [Dubois & Prade, 88]) is a particular numerical formalism based on the use of the so-called *possibility* and *necessity* measures that provide to what extent a crisp piece of knowledge can be considered plausible and certain respectively. Even if formulas bear numerical possibilities, we may be interested not in the values themselves but only in their comparison, i.e. in formulas such as $A \triangleleft B$ saying for instance that $B$ is at least as possible as $A$. In this direction, a qualitative characterization of possibility measures was first given in [Dubois 86], where it was established a formal relation between possibility theory and qualitative possibility relations. Since then several logics of qualitative possibility have been proposed to allow specifying partial qualitative possibility relations and deriving consequences based on this partial information. Fariñas and Herzig [Fariñas & Herzig, 91] axiomatize the logic of qualitative possibility called $QPL$, introducing a conditional connective $\geq$, and they relate it to the Lewis' conditional logic $VN$. They also propose a multi-modal logic based on sphere semantics to fully support possibility theory. Boutilier presents in [Boutilier, 92] a possible worlds semantics for qualitative possibility relations and defines two modal logics, $CO$ and $CO^*$, and he makes use of two modalities to capture possibilistic logic. One corresponds as usual to truth at accessible worlds while the other to truth to inaccessible worlds. Finally, in [Bendová & Hájek, 93] qualitative possibilistic logic is related to a tense logic with finite linearly preordered time called $FLPOT$, pushing further some aspects of the previous works, specially the incompleteness of $QPL$ and the precise relation of the above modal systems with unary modalities to $QPL$. It is also worth noticing that similar attempts of relating other quantitative and qualitative uncertainty measures have been proposed in the literature. See for instance [Segerberg, 71] or [Bacchus, 90] for the case of probability measures or [Harmanec & Hájek, 94], [Resconi, Klir & St. Clair, 92] for belief and plausibility measures. See also [Wong et al., 91].

On the other hand, vagueness (fuzziness) concerns degrees of truth, usually numerical, and therefore one is led to many-valued logics as underlying formalism. In this sense, fuzzy logic deals with fuzzy propositions that may have intermediate degrees of truth and it is usually understood as truth-functional, i.e. the truth degree of a compound formula is a function of the truth degrees of its subformulas using suitable connectives. Because of this key feature, fuzzy logic departs from any uncertainty logic.

So far, as argued elsewhere ([Godo & Mantaras, 93], [Hájek, 93], [Hájek & Harmancová, 93]), we have seen that qualitative uncertainty (possibility) leads to modal logics, while fuzziness leads to many-valued logics. Therefore, presence of both fuzziness and uncertainty leads to calculi that are both many-valued and modal. Modal many-valued logics have been already investigated by some authors, see for instance [Ostermann, 88] or [Fitting, 92]. Our goal is to study logical systems of qualitative possibility dealing with fuzzy propositions. It is clear then that the first question one is faced with is how to compare formulas of fuzzy logic with respect to their possibility or, more general, with respect to the possibility measures of worlds on which they have a given truth-value. This question has not an unique answer. For instance, in a recent and very related works [Hájek & Harmancová, 93], [Hájek, Harmancová & Verbrugge, $\infty$] the possibility of a fuzzy formula is described by a fuzzy truth-value, i.e. a function that provides, for each truth-value $\alpha$, the possibility of the fuzzy formula being $\alpha$-true. The corresponding modal calculus $QFL$ of comparison of possibilities is there shown to have a faithful interpretation in a certain many-valued tense logic (with linearly preordered time), and thus naturally extending the results in [Bendová & Hájek, 93].

In the present paper we investigate qualitative fuzzy logics based on an alternative definition of possibility of a fuzzy formula, following the well known Zadeh's extension of classical possibility measures to fuzzy propositions (cf. [Zadeh, 78] [Dubois & Prade, 85]). With this definition, the possibility of a formula of fuzzy logic is again only one value. Furthermore, in our setting the possibility of a formula is just a truth value. To simplify matters, we make the following design choices: we investigate finitely valued Łukasiewicz fuzzy logics where the set $Values$ of truth values coincides with the set of allowed possibility degrees. We show that this notion is useful to express statements about qualitative possibility comparisons. The corresponding comparative logic $QFL_2$[1] is shown to satisfy reasonable axioms and to be closely related to the many-valued analogue $MVS5$ of the modal logic $S5$. In Artificial Intelligence, $S5$ is often understood as the logic of knowledge whereas its weakening $KD45$ is understood as the logic of belief (cf [Voorbraak, 93], [Halpern & Moses, 92]). We relate our calculi also to $KD45$ and its many-valued counterpart MVKD45. Completeness theorems for $MVS5$ and $MVKD45$ are obtained. This is not a too surprising result; [Fitting, 92] has much about completeness of many valued modal logics and [Ostermann, 88] too. But we present Hilbert-style systems and get their completeness. Cf. also [Nakamura, 91a, 91b] for a different approach.

The paper is organized as follows. In Section 2 the underlying many-valued propositional calculus is described. In Section 3, possible world semantics for the qualitative fuzzy logic $QFL_2$ is introduced, while in Section 4 we prove its faithful interpretation on the many-valued modal $MVS5$, also described in this section. Finally, in Section 5 the logic $MVKD45$ is presented and proved that $QFL_2$ can be also faithfully interpreted there. We end with some concluding remarks.

## 2 MANY-VALUED PROPOSITIONAL CALCULUS USED

Fix a natural number $n \geq 2$; we base our investigations over the n-valued propositional Łukasiewicz's logic $L_n$, as described e.g. in [Gottwald, 88]. This choice does not imply that we claim Łukasiewicz's logic be the only formal logical system for fuzzy logic. Our set of truth values is $Values = \{0, 1/(n-1), \ldots, 1\}$. Principal connectives are implication and negation, denoted respectively by $\rightarrow$ and $\neg$, with Łukasiewicz semantics: $\neg A$ is interpreted by $1 - x$ and $A \rightarrow B$ by $min(1 - x + y, 1)$ where $x, y \in Values$ are the interpretations of $A$ and $B$ respectively; occasionally we write $I(x, y)$ for $min(1 - x + y, 1)$. Other connectives are defined from $\rightarrow$ and $\neg$, in particular, there are two conjunctions and two disjunctions. We have the connectives $\wedge$ and $\vee$, interpreted by $min(x, y)$ and $max(x, y)$ respectively, and $\&$ interpreted by $max(0, x + y - 1)$, and its dual $\underline{\vee}$ interpreted by $min(1, x + y)$. The equivalence connective $\leftrightarrow$ is interpreted by $min(I(x, y), I(y, x))$. A complete axiomatization of $L_n$ extends the celebrated axiomatization of $L_\infty$ [Rose and Rosser, 58] by some few additional axioms, namely:

- $A \rightarrow (B \rightarrow A)$,
- $(A \rightarrow B) \rightarrow ((B \rightarrow C) \rightarrow (A \rightarrow C))$,
- $(\neg B \rightarrow \neg A) \rightarrow (A \rightarrow B)$,
- $((A \rightarrow B) \rightarrow B) \rightarrow ((B \rightarrow A) \rightarrow A)$,
- $\sum_{i=1}^{n} A \rightarrow \sum_{i=1}^{n-1} A$,
- $\sum_{i=1}^{n-1} (\prod_{j=1}^{m} A \underline{\vee} (\neg A \& \sum_{j=1}^{m-1} A))$,

for $1 < m < n - 1$ such that $m - 1$ does not divide $n - 1$, and where $\sum_{i=1}^{n} A$ stands for $A\underline{\vee}\ldots^{n)}\ldots\underline{\vee}A$ and $\prod_{i=1}^{n} A$ for $A\&\ldots^{n)}\ldots\&A$. Note the 1-tautologies $(A \rightarrow (B \rightarrow C)) \leftrightarrow ((A\&B) \rightarrow C)$ and $(\neg(A \rightarrow B)) \leftrightarrow (A\&\neg B)$. Further note the 1-tautology $(A \rightarrow B) \leftrightarrow (\neg B \rightarrow \neg A)$. We extend $L_n$ by unary connectives (coefficients) $(i)$ for each $i \in Values$; the value of $(i)A$ is 1 iff the value of $A$ is $i$, otherwise the value of $(i)A$ is 0. In fact the connectives $(i)$ are definable in $L_n$, see [Ostermann, 88] or [Gottwald, 88]. Finally notice that Dienes and Gödel's implications are also definable in $L_n$ as $\neg A \vee B$ and $(1)(A \rightarrow B) \vee B$ respectively.

In the sequel, the constant $True$ will stand for $A \rightarrow A$, $False$ for $\neg True$, and $B_0$-formulas will be formulas generated from formulas of the form $(i)A$ using connectives and coefficients. Clearly, each such a formula is Boolean in the semantic sense: it takes only values 1 and 0. (Later we introduce another class of formulas that are Boolean). We list next some axioms on formulas containing coefficients (cf. [Hájek & Harmancová, 93]); in fact they are provable (by definability of coefficients and completeness of axioms for $L_n$).

---

[1] The subindex 2 distinguishes our logic from the logic $QFL$ of [Hájek & Harmancová].

$$\bigvee_i (i)A, \quad \bigwedge_{i \neq j} \neg((i)A \wedge (j)A),$$

$(i)A \to (1-i)\neg A$

$((i)A \wedge (j)B) \to (t_*(i,j))(A * B),$

for $*$ being $\wedge, \vee, \&, \underline{\vee}, \to, \leftrightarrow$ and $t_*$ being the corresponding truth interpretation,

$(1)A \to A,$

for $B_0$ - formulas $A, B, C$:

$A \leftrightarrow (1)A,$
$(A \to (B \to C)) \to ((A \to B) \to (A \to C))$[2]

The only deduction rule is *modus ponens*. This ends our description of the underlying propositional calculus.

## 3 KRIPKE MODELS, POSSIBILITIES

Our possibilistic Kripke models over a set *Atom* of propositional atoms have the form

$$K = \langle W, \Vdash, \pi \rangle$$

where $W$ is a non-empty set of possible worlds, $\Vdash$ maps $Atom \times W$ into $Values$, $\pi$ maps $W$ into $Values$ and $max_{w \in W} \pi(w) = 1$. Recall that we consider possibilities taking values only in $Values$. We extend $\Vdash$ in the obvious way to a mapping (denoted again by $\Vdash$) of $Form_0 \times W$ into $Values$ where $Form_0$ is the set of all formulas of our propositional logic satisfying the usual inductive conditions. We write $\| A \|_w = i$ for $\Vdash (A, w) = i$. Concerning satisfiability, we shall write $w \Vdash A$ iff $\| A \|_w = 1$. (Note $\| A \|_w = i$ iff $w \Vdash (i)A$.) The corresponding notions of validity and semantical entailment are the usual ones.

After Zadeh (cf. [Zadeh 1978], [Dubois & Prade, 86]), we introduce the following notion of the *possibility* degree of a (fuzzy) formula $A \in Form_0$ in a possibilistic Kripke model, that extends to many-valued propositions the notion of classical possibility measure for two-valued propositions which plays a central role in possibilistic logic.

---

[2]Note that this formula is one of famous axioms of the classical two-valued propositional calculus; it is sound for boolean formulas but *not* for all formulas.

**Definition 3.1** $\Pi(A) = sup_w(\pi(w) \wedge \| A \|_w)$.

The corresponding dual notion of *necessity* can be then as $N(A) = 1 - \Pi(\neg A) = inf_w(1 - \pi(w) \vee \| A \|_w)$. The idea behind the above definition is to use it in next sections to interpret in our comparative logic $QFL_2$ sentences of type *B is at least as possible as A*, being $A$ and $B$ many-valued, as $\Pi(A) \leq \Pi(B)$. This interpretation extends to the fuzzy (many-valued) case, in a different way than $QFL$, the comparison of possibilities that is present in the qualitative possibilistic logics $QPL, CO$ and $FLPOT$ mentioned in the introduction section. However, it is worth noticing that other ways of extending the notion of possibility for fuzzy propositions have been also advocated; see [Dubois & Prade, 92] for a discussion of such extensions. Next lemmas show a characterization of possibility measures $\Pi$ given by possibilistic Kripke models.

**Lemma 3.2** *(cf. [Dubois & Prade, 1988]) For each possibilistic Kripke model $K$, $\Pi$ satisfies the following:*

($\Pi$1) $\Pi(True) = 1$, $\quad \Pi(False) = 0$

($\Pi$2) $\Pi(A \vee B) = max(\Pi(A), \Pi(B))$

($\Pi$3) if $L_n \vdash A \leftrightarrow B$ then $\Pi(A) = \Pi(B)$

($\Pi$4) $\Pi(A) = \bigvee_i (i \wedge \Pi((i)A))$

*Proof:* We only prove the last equality. $\Pi(A) = sup_w(\pi(w) \wedge \| A \|_w) = sup_i((sup_{\|A\|_w = i} \pi(w)) \wedge i) = sup_i(i \wedge \Pi((i)A))$. ∎

**Lemma 3.3** *Assume the set Atom of propositional atoms to be finite. If a mapping $\Pi : Form_0 \to [0,1]$ satisfies the axioms ($\Pi 1 - \Pi 4$) of the preceding lemma then there is a finite Kripke model $K$ such that $\Pi$ is the possibility given by $K$.*

*Proof:* Note that ($\Pi$4) guarantees that it is enough to produce a Kripke model $K$ whose possibility coincides with $\Pi$ for $B_0$-formulas. Now it is easily seen that each $B_0$-formula is equivalent to a Boolean combination of formulas of the form $(i)p$ where $p$ is a propositional atom. Therefore, we may produce a model $K$ in full analogy to the two-valued case: each $B_0$-formula $B$ is $L_n$-equivalent to a disjunction of maximal elementary conjunctions of the form $\bigwedge_{i=1}^{m}(j_i)p_i$, where $m$ is the cardinality of the set *Atom*, thus for each such $B$ there is a maximal elementary conjunction $C$ such that $\Pi(B) = \Pi(C)$. Thus we construct our model from the elementary conjunctions in the usual way. ∎

Notice that, given a possibilistic Kripke model $K = \langle W, \Vdash, \pi \rangle$ over a set *Atom* of atoms, if $p$ is a propositional variable not in *Atom*, $K$ clearly determines uniquely a model $K' = \langle W, \Vdash' \rangle$ over $Atom' = Atom \cup \{p\}$ such that $\Vdash'$ coincides with $\Vdash$ on *Atom* and $\Vdash'(p, w) = \pi(w)$ for each $w \in W$. ($K'$ has no explicit structure on $W$.) Models of the form $\langle W, \Vdash \rangle$ will be called MVS5-*models* (for obvious reasons, see below). If $K' = \langle W, \Vdash' \rangle$ is an MVS5-model over $Atom \cup \{p\}$ such that $max\{\| p \|_w | w\} = 1$ then it determines uniquely a possibilistic model $K = \langle W, \Vdash, \pi \rangle$ where $\Vdash$ is the restriction of $\Vdash'$ to $Atom \times W$ and $\pi(w) = \Vdash'(p, w)$ for all $w \in W$. We shall identify $K$ and $K'$ without any danger of misunderstanding.

In the next section we shall introduce various modalities defined in our Kripke models, among them the modality of comparison of possibilities.

## 4   SOME MODALITIES, THE FUZZY LOGIC $QFL_2$ AND ITS RELATION TO THE MODAL MANY-VALUED LOGIC $MVS5$

We enrich our language by three modalities, $\Diamond, \Diamond_p$ and $\lhd_p$ (for $Atom' = Atom \cup \{p\}$) and define their semantics as follows.

$$\| \Diamond A \|_w = max\{\| A \|_{w'} |\ w' \in W\};$$
$$\| \Diamond_p A \|_w = max\{\| p \wedge A \|_{w'} |\ w' \in W\};$$
$$A \lhd_p B\ is\ \Diamond_p A \to \Diamond_p B$$

the corresponding duals $\Box A, \Box_p A$ and $A \prec_p B$ being defined as $\neg \Diamond \neg A, \neg \Diamond_p \neg A$ and $\neg(\neg A \lhd_p \neg B)$ respectively.

Note that formulas $\Diamond A, \Diamond_p A$ and $A \lhd_p B$ take a *constant* value independently of a given $w \in W$. Thus we shall write from now on $\| \Diamond A \|, \| \Diamond_p A \|$ etc. Next lemma summarizes properties and relations among the above modalities; note that we work with models $K'$ in which $max(\| p \|_w | w) = 1$, i.e. $\| \Diamond p \| = 1$.

**Lemma 4.1** *The following properties hold for any formulas $A$ and $B$:*

*(a)* $\| \Box A \| = min\{\| A \|_w | w \in W\}$

*(b)* $\| \Diamond_p A \| = \Pi(A)$,
    $\| \Box_p A \| = N(A)$

*(c)* $\Diamond_p A$ *is equivalent to* $\Diamond(p \wedge A)$,
    $\Box_p A$ *is equivalent to* $\Box(\neg p \vee A)$

*(d)* $A \prec_p B$ *is equivalent to* $\Box_p A \to \Box_p B$

*(e)* $\| A \lhd_p B \| = 1$ *iff* $\Pi(A) \le \Pi(B)$,
    $\| A \prec_p B \| = 1$ *iff* $N(A) \le N(B)$

*(f)* $\Diamond_p A$ *is equivalent to* $True \lhd_p A$,
    $\Box_p A$ *is equivalent to* $True \prec_p A$

where $\Pi$ *(and $N$) is defined by extension of definition 3.1 to any formula in the obvious way.*

Thus it is clear that the modality $\Diamond_p$ captures the possibility $\Pi$ whereas the binary modality $\lhd_p$ captures the comparison of possibilities. Furthermore, the lemma shows that $\Diamond_p$ and $\lhd_p$ are interdefinable and both are definable from $\Diamond$. So it makes sense, for our purposes of investigating the logic of comparison of possibilities, to axiomatize the $\Box$ modality. This is our next task.

First, observe that if a formula $A$ is Boolean (takes only values 0,1) then $\Diamond A$ is also Boolean; but of course $\Diamond_p A$ need not be Boolean. Define the class of *B-formulas* to be the class of formulas resulting from formulas of the form $(i)A$, where $A$ is an arbitrary formula, possibly containing modalities, using connectives. Thus each $B$-formula is Boolean.

**Definition 4.2** *The modal logic MVS5 has the following axioms:*

- *propositional axioms as in Section 2 but the axioms*

  $A \leftrightarrow (1)A$,
  $A \to (B \to C) \to ((A \to B) \to (A \to C))$

  *are postulated for all B-formulas, not only $B_0$-formulas;*

- *modal axioms:*

  $\left.\begin{array}{l} \Box(A \to B) \to (\Box A \to \Box B) \\ \Box A \to A \\ \Box A \to \Box\Box A \\ \Diamond A \to \Box\Diamond A \end{array}\right\}$ this is S5

  $(\ge i)(\Box A) \leftrightarrow \Box(\ge i)A$ } Fitting – like

  *(Clearly, $(\ge i)A$ stands for $\bigvee_{j \ge i}(j)A$ .)*

*Deduction rules are : modus ponens, necessitation and "from $A$ infer $(1)A$".*

**Theorem 4.3** *(Completeness). $MVS5 \vdash A$ iff $A$ is a 1-tautology for MVS5-models.*

The proof is standard and sketched at the end of this section.

Next step is to formally introduce our qualitative modal logic for comparison of possibilities of fuzzy propositions, and to faithfully embed it in MVS5

**Definition 4.4** *The qualitative modal fuzzy logic $QFL_2$ over a set Atom of atoms has formulas built up from atoms (propositional variables) using logical connectives and a binary modality $\lhd$; models are possibilistic models $\langle W, \Vdash, \pi \rangle$ and the semantics is*

$$\| A \lhd B \| =$$
$$= I(max_w(\| A \|_w \wedge \pi(w)), max_w(\| B \|_w \wedge \pi(w))).$$

It is worth noticing that $\| A \lhd B \| = 1$ iff $\Pi(A) \le \Pi(B)$.

**Definition 4.5** *Given a variable $p$ not in Atom, the translation of $QFL_2$ formulas to MVS5 formulas is defined recursively as follows:*

$q^*$ is $q$ for $q \in Atom$,

$(A \to B)^*$ is $A^* \to B^*$ and similarly for other connectives,

$(A \lhd B)^*$ is $A^* \lhd_p B^*$.

Finally define $A^{**}$ to be $(1)\Diamond p \to A^*$.

**Theorem 4.6** *The above mapping ** is a faithful interpretation of $QFL_2$ in MVS5, i.e. a formula $A$ of $QFL_2$ is a 1-tautology iff $A^{**}$ is a MVS5-tautology.*

*Proof:* Let $K = \langle W, \Vdash, \pi \rangle$ be a model of $QFL_2$ and let $w\Vdash(i)A$. Then $K' = \langle W, \Vdash' \rangle$ constructed above is a model of MVS5 and $K'\Vdash'(1)\Diamond p, w\Vdash'(i)A^*$ and hence $w\Vdash' A^{**}$. Thus if $A^{**}$ is a 1-tautology of MVS5, then $A$ is a 1-tautology of $QFL_2$.

Conversely, let $K_1 = \langle W, \Vdash' \rangle$ be a MVS5-model (of the language extended by $p$). If $K_1\Vdash'(<1)\Diamond p$ then clearly $K_1\Vdash'(1)A^{**}$; if $K_1\Vdash'(1)\Diamond p$ then $K_1 = K'$ for the obvious $K$. If $w\Vdash'(i)A^{**}$ then $w\Vdash'(i)A^*$ and $w\Vdash(i)A$. Thus if $A$ is a 1-tautology of $QFL_2$ then $A^{**}$ is a 1-tautology of MVS5. ∎

In the rest of this section we sketch a proof of the completeness of MVS5.

Recall B-formulas; a *theory* is a set of B-formulas including all formulas $(1)C$ where $C$ is MVS5-provable. $T \vdash C$ ($T$ proves $C$) if there is a proof of $C$ from $T$ using only modus ponens (no necessitation). $T$ is *complete* if for each B-formula $C$, $T \vdash C$ or $T \vdash \neg C$. As usual, it suffices to show the following Lemma.

**Lemma 4.7** *(Main Lemma) If $T_0$ is a complete theory and $T_0 \vdash (i)A$ then there is a model $K = \langle W, \Vdash \rangle$ and a $w \in W$ such that $w\Vdash(i)A$.*

To build such a model we need first some previous results.

**Definition 4.8** *Let $T$ and $T_0$ be complete theories. We say that $T$ and $T_0$ are equivalent, written $T \approx T_0$, provided that for each $i$ and $A$, $T \vdash (i)\Box A$ iff $T_0 \vdash (i)\Box A$, or equivalently, $T \vdash (i)\Diamond A$ iff $T_0 \vdash (i)\Diamond A$.*

**Lemma 4.9** $MVS5 \vdash (i)\Diamond C \to (\Box(\leq i)C \land \Diamond(i)C)$.

**Corollary 4.10** *If $T_0 \vdash (i)\Diamond C$ then*

(a) $(\forall T \approx T_0)(T \vdash (\leq i)C)$

(b) $(\exists T \approx T_0)(T \vdash (i)C)$

*Proof:* Easy, from compactness, like in [Hájek & Harmancová, 93]. ∎

Now the definition of the model follows.

**Definition 4.11** *For each $i$ and $C$ such that $T_0 \vdash (i)\Diamond C$, let $T_C$ be a complete theory satisfying (b) in the above corollary. We define the model $K = \langle W, \Vdash \rangle$ such that the set of models is $W = \{T_0\} \cup \{T_C \mid C \text{ arbitrary}\}$ and the forcing relation is defined by $T\Vdash(i)p$ iff $T \vdash (i)p$, for any $T \in W$.*

Finally, completeness comes immediately from next lemma.

**Lemma 4.12** *For each formula $B$, and each $T \in W$, $T \vdash (i)B$ iff $T\Vdash(i)B$.*

*Proof:* Induction step for $\Diamond B$: Assume $T \vdash (i)\Diamond B, T \in W$. Then $T \vdash \Box(\leq i)B$, thus for each $T' \in W$, $T' \vdash \Box(\leq i)B$, thus $T' \vdash (\leq i)B$ (using $\Box D \to D$), thus $T'\Vdash(j)B$ for some $j \leq i$, by the induction hypothesis. On the other hand, $T_0 \vdash \Diamond(i)B$ implies $T_B \vdash (i)B$ and by the induction hypothesis, $T_B\Vdash(i)B$. Hence $i = max\{\| B \|_T \mid T \in W\}$, $i = \| \Diamond B \|$. This completes the proof of this lemma and of the Main Lemma 4.7. ∎

**Corollary 4.13** *MVS5 is complete with respect to the given semantics.*

## 5 THE LOGIC $QFL_2$ AND A MANY-VALUED BELIEF LOGIC

The $QFL_2$ comparative modality $\lhd$ introduced in the previous section relies fundamentally on the MVS5 modality $\Diamond_p$. Therefore it seems interesting to investigate a possible axiomatization of the modality $\Diamond_p$ itself, without needing to refer it to any other modality. To this end, in this section we relate our $QFL_2$ to a many-valued version of the belief logic $KD45$ (see e.g. [Voorbraak]). Our $MVKD45$ will be a subtheory of $MVS5$ (like $KD45$ is a subtheory of $S5$); if there are only two values ($Values = \{0,1\}$) then $MVKD45$ becomes $KD45$ like $MVS5$ becomes $S5$. Moreover, a faithful embedding of $QFL_2$ into $MVKD45$ is very easy to define.

**Models** of $MVKD45$ are again possibilistic Kripke structures $K = \langle W, \Vdash, \pi \rangle$, where $\pi$ is a normalized possibility distribution on $W$ with values in $Values$ that can be understood as a many-valued accessibility relation $R$ defined as $R(w, w') = \pi(w')$. Such many-valued accessibility relations already occur in [Fitting, 92].

The semantics of the $MVKD45$ modalities $\Box, \Diamond$ is as follows:

$$\|\Diamond B\| = max_w(\| B \|_w \land \pi(w)), \quad \|\Box B\| = \|\neg\Diamond\neg B\|.$$

Next lemmas show the $MVKD45$-validity of some formulas that will be taken later as axioms of our logic.

**Lemma 5.1** *The formulas*

$$\Box(A \to B) \to (\Box A \to \Box B),$$

$\Box A \leftrightarrow \Box\Box A$, $(j)\Box A \leftrightarrow \Box(j)\Box A$,

$\Diamond A \leftrightarrow \Box\Diamond A$, $(j)\Diamond A \leftrightarrow \Box(j)\Diamond A$ and $(1)\Diamond True$

are 1-tautologies.

*Proof:* We only prove the first formula (axiom K). The rest are easily proved by straightforward computations. It suffices to show that

$$\| \Box A \to \Box B \| \geq \| \Box(A \to B) \|.$$

We have in the one hand $\| \Box A \to \Box B \| = I(min_w\{1 - \pi(w) \vee \|A\|_w\}, min_{w'}\{1 - \pi(w') \vee \|B\|_{w'}\}) = max_w min_{w'} I(1 - \pi(w) \vee \|A\|_w, 1 - \pi(w') \vee \|B\|_{w'}) \geq min_w I(1 - \pi(w) \vee \|A\|_w, 1 - \pi(w) \vee \|B\|_w)$, and in the other hand $\| \Box(A \to B) \| = min_w\{1 - \pi(w) \vee I(\|A\|_w, \|B\|_w)\}$. Thus if we prove

$$I(1 - \pi(w) \vee \|A\|_w, 1 - \pi(w) \vee \|B\|_w) \geq$$
$$\geq 1 - \pi(w) \vee I(\|A\|_w, \|B\|_w)$$

the lemma will be proved. But observe that:

- $I(1 - \pi(w) \vee \|A\|_w, 1 - \pi(w) \vee \|B\|_w) = min(I(1-\pi(w), 1-\pi(w) \vee \|B\|_w), I(\|A\|_w, 1-\pi(w) \vee \|B\|_w) = I(\|A\|_w, 1-\pi(w) \vee \|B\|_w) \geq I(\|A\|_w, \|B\|_w)$

- $I(1-\pi(w) \vee \|A\|_w, 1-\pi(w) \vee \|B\|_w) \geq 1 - \pi(w) \vee \|B\|_w \geq 1 - \pi(w)$

Thus the above inequality holds and the lemma too. ∎

Recall the notion of a maximal elementary conjunction (m.e.c.) of the form $\bigwedge_{i=1}^m (j_i)p_i$. Let $K = (W, \Vdash, \pi)$ be a possibilistic model and let $E$ be a m.e.c. and $A$ any formula. Then we have the following further lemmas.

**Lemma 5.2** *There exists a unique $j \in Values$ such that for each $w \in W$, $w \Vdash E \to (j)A$.*

*Proof:* Evident for $A$ atomic, induction step clear for connectives as well as for modalities (since the truth value of a modalized formula is independent of $w$). ∎

**Lemma 5.3** $K \Vdash (>0)\Diamond((j)A \wedge E) \to (E \to (j)A)$

*Proof:* If $K \Vdash (>0)\Diamond((j)A \wedge E)$ then there is a $w \in W$ such that $\pi(w) \wedge \|(j)A \wedge E\|_w > 0$, thus $\|(j)A \wedge E\|_w = 1$, $w \Vdash (j)A \wedge E$, so this $j$ is the unique $j$ of Lemma 5.2. Thus, for each $w_0 \in W, w_0 \Vdash E \to (j)A$. ∎

**Lemma 5.4** $K \Vdash ((i)\Diamond A \wedge E) \to (\leq i)(A \wedge \Diamond E)$

*Proof:* Assume $w \Vdash (i)\Diamond A \wedge E$; then $\| \Diamond E \| = max\{\pi(v) \mid v \Vdash E\} = \pi(v_0)$, and $i = max_v(\pi(v) \wedge \|A\|_v)$. Let $j$ be such that $(\forall v)(v \Vdash E \to (j)A)$; we have $\|A \wedge \Diamond E\|_w = \|A \wedge \Diamond E\|_{v_0}$ (since $\|A\|_w = \|A\|_{v_0} = j$) and $\|A \wedge \Diamond E\|_{v_0} = \|A\|_{v_0} \wedge \pi(v_0) \leq \|\Diamond A\| = i$. Thus $\|A \wedge \Diamond E\|_w \leq i$. ∎

**Lemma 5.5** $K \Vdash (i)\Diamond A \to \bigvee_E (\geq i)\Diamond(E \wedge (i)(A \wedge \Diamond E))$.

*Proof:* If $i = 0$ it is obvious. If $i = max_v(\|A\|_v \wedge \pi(v)) > 0$ then, it is easy to prove that, there exists $v_0$ and $E$ such that $\|A\|_{v_0} \wedge \pi(v_0) = i$, $v_0 \Vdash E$ and $\pi(v_0) = \| \Diamond E \|$. Therefore $\|A\|_{v_0} \wedge \pi(v_0) = \|A\|_{v_0} \wedge \|\Diamond E\| = i$. Thus $\|E \wedge (i)(A \wedge \Diamond E)\|_{v_0} = 1$, $\|E \wedge (i)(A \wedge \Diamond E)\|_{v_0} \wedge \pi(v_0) = i$, and therefore $\| \Diamond (E \wedge (i)(A \wedge \Diamond E)) \| \geq i$. ∎

Now we are ready to present our **axioms of MVKD45**.

**Definition 5.6** *The modal logic MVKD45 has the following axioms:*

- *axioms of propositional calculus (as above)*

- *axioms of KD45:*

  $\Box(A \to B) \to (\Box A \to \Box B)$,

  $\Box A \leftrightarrow \Box\Box A$, $\Diamond A \leftrightarrow \Box\Diamond A$

- $(j)\Box A \leftrightarrow \Box(j)\Box A$, $(j)\Diamond A \leftrightarrow \Box(j)\Diamond A$,

- $(1)\Diamond True$

- $((j)\Diamond A \wedge E) \to (\leq j)(A \wedge \Diamond E)$,    (1)
  *being E a m.e.c.*

- $(j)\Diamond A \to \bigvee_E (>0)\Diamond(E \wedge (j)(A \wedge \Diamond E))$,    (2)
  *for $j > 0$ and E being a m.e.c.*

*Deduction rules are Modus Ponens, necessitation and "from A infer (1)A".*

Previous lemmas 5.1 to 5.5 prove the soundness of $MVKD45$. Therefore, the rest of this section is devoted to get the completeness results for our logic as well as the embedding of $QFL_2$ into $MVKD45$ as mentioned before. The techniques are similar to the case of $MVS5$, i.e. for any formula provable in a complete theory we can build a possibilistic model where it is satisfiable. First of all we need the following lemma.

**Lemma 5.7** *Let $T_0$ be a complete theory such that $T_0 \vdash (i)\Diamond C$. Then*

(a) $(\forall T \approx T_0)(\forall E)(T \vdash (E \to (\leq i)(C \wedge \Diamond E)))$,

(b) $(\exists T \approx T_0)(\exists E)(T \vdash (E \wedge (i)(C \wedge \Diamond E)))$.

*Proof:* (b - sketch) Assume $T_0 \vdash (j)\Diamond C$, $j > 0$. Then for some $E$, $T_0 \vdash (>0)\Diamond(E \wedge (j)(C \wedge \Diamond E))$. Put $D = (E \wedge (j)(C \wedge \Diamond E))$. Let $T_0 \vdash (i_k)\Box B_k$, $k = 1, \ldots, n$; then $T_0 \vdash \Box(i_k)\Box B_k$ and $T_0 \vdash \Box \bigwedge_{k=1}^n (i_k)\Box B_k$ (note that $(i_k)\Box B_k$ is a B-formula!). Since MVKD45 proves $\Diamond D \to (\Box H \to \Diamond(D \wedge H))$ for $H, D$ being B-formulas, we get, for $H = \bigwedge(i_k)\Box B_k$, $T_0 \vdash \Diamond D \to \Diamond(D \wedge H)$, thus $T_0 \vdash (>0)\Diamond D \to (>0)\Diamond(D \wedge H)$, thus $T_0 \vdash (>0)\Diamond(D \wedge H)$, and therefore $D \wedge H$ is consistent[3].

---
[3] Recall that a formula $A$ is $MVKD45$-consistent if $MVKD45 \not\vdash \neg A$.

Consequently, $D$ is consistent with the set of all $T_0$-provable formulas of the form $(i)\Box B$, completing this theory we get our $T$. ∎

This last lemma enables us to define our model as follows.

**Definition 5.8** *For each $i$ and $C$ such that $T_0 \vdash (i)\Diamond C$ let $T_C$ be a theory $T$ as in (b) in the above lemma. We define the model $(W, \Vdash, \pi)$ such that the set of worlds is $W = \{T_0\} \cup \{T_C \mid C\}$; for $T \in W$, the forcing relation is defined by $T\Vdash(i)p$ iff $T \vdash (i)p$, and finally, the possibility distribution is given by $\pi(T) = i$ iff $T \vdash E \wedge (i)\Diamond E$.*

Completeness is obtained by proving next main lemma in a similar way as in lemma 4.12.

**Lemma 5.9** *For each $i, B$, and for each $T \in W$,*
$$T\Vdash(i)B \quad \text{iff} \quad T \vdash (i)B.$$

**Corollary 5.10 (Completeness)** *MVKD45 is complete with respect to the given semantics.*

Finally, $QFL_2$ is related to MVKD45 in the way next theorem shows.

**Theorem 5.11** *Define an interpretation of $QFL_2$-formulas in MVKD45 by putting $(A \triangleleft B)^* = (\Diamond A \to \Diamond B)$ and extending trivially to all $QFL_2$-formulas. Then $*$ is a faithful interpretation of $QFL_2$ in MKVD45, i.e. a $QFL_2$-formula $A$ is a $QFL_2$-tautology iff $A^*$ is a MVKD45-tautology.*

## 6 CONCLUDING REMARKS AND FUTURE WORK

In this paper we have investigated, from a logical point of view, a modality for comparison of possibilities of fuzzy propositions. In this sense, this paper tackles the same problem as in [Hájek & Harmancová, 93] but with another approach. Taking as reference the Zadeh's extension of the concept of possibility measures to fuzzy propositions, the corresponding comparative logic $QFL_2$ has been related to two many-valued modal systems namely $MVS5$ and $MVKD45$, for which complete axiom systems, Hilbert style, have been given. However a number of open questions remain for future investigation. Some of them are listed below.

(1) It would be desirable to replace axioms (1), (2) of definition 5.6 by some other more elegant axioms (in particular not dealing explicitly with m.e.c.'s). One tautology more similar to the axioms of MVS5 is

$$(\geq i)\Box C \leftrightarrow (\geq i)\Box(\geq i)C.$$

(2) Our choice of semantics of $\Box A$ in MVKD45 might seem unnatural:
$\|\Box A\| = min_w((1-\pi(w)) \vee \|A\|_w)$. A seemingly more natural choice would be $\|\Box A\| = min_w(\pi(w) \to \|A\|_w)$ but then to have duality of $\Box$ and $\Diamond$ we should have $\|\Diamond A\| = max_w(\pi \;\&\; \|A\|_w)$ (strict conjunction). Unfortunately, for this semantics of $\Box$ one of the basic axioms of modal logic, namely $\Box(A \to B) \to (\Box A \to \Box B)$, is not a tautology.

(3) To find an elegant (non-pedestrian) axiomatization of $QFL_2$ in its own language still remains. As a matter of fact, it is worth noticing that some of the axioms of Fariñas and Herzig's $QPL$ logic, e.g.

- $(A \triangleleft B) \vee (B \triangleleft A)$
- $(A \triangleleft B) \to ((A \vee C) \triangleleft (B \vee C))$
- $A \triangleleft True$
- $\neg(True \triangleleft False)$

are 1-tautologies of $QFL_2$ too, so they are potential candidates.

(4) Furthermore, one should give up the assumption that $Values$ is finite and study the full Lukasiewicz's logic with real values.

(5) The relationship between the semantics of the comparative modality in [Hájek & Harmancová, 93] and in this present paper it is also a matter of future research.

## Acknowledgements

Francesc Esteva, Pere Garcia and Lluis Godo have been partially supported by the Esprit III Basic Reserach Action 6156 DRUMS II and by a grant from the Spanish DGICYT (Pr. No. PB91-0334-C03-03).

Petr Hájek and Dagmar Harmancová have been partially supported by the grant Nr. 130108 of the Grant Agency of the Academy of Sciences of Czech Republic.

Partial support by the COPERNICUS grant No.10053 (MUM) in the final stages of preparation of this paper is also acknowledged.